\documentstyle[prb,aps,twocolumn]{revtex}
\input epsf
\epsfverbosetrue
\begin{document}
\draft
\flushbottom
\twocolumn[
\hsize\textwidth\columnwidth\hsize\csname @twocolumnfalse\endcsname

\title{Spectral Function Analysis of Fermi Liquids and of Composite Fermions in
a Finite Magnetic Field: Renormalised Gaps} 

\vskip .5in

\author{S. Curnoe$^1$ and P. C. E. Stamp$^{1,2}$}


\address{$^1$Department of Physics \& Astronomy
      and  $^2$Canadian Institute for Advanced Research\\
 University of British Columbia,
Vancouver, B.C. V6T 1Z1, Canada}
\maketitle
\maketitle
\tightenlines
\widetext
\advance\leftskip by 57pt
\advance\rightskip by 57pt

\begin{abstract}
We consider the self-energy and quasiparticle spectrum, for both 
electrons interacting with phonons, and composite fermions 
interacting with gauge fluctuations.  In both cases we incorporate
the singular structure arising from Landau level quantization in a 
finite field.   This is then used to determine the renormalised gap
between the Fermi energy and the first excited states.  

The electron-phonon problem is treated for both Debye and Einstein
phonons.  In the case of composite fermions, it is found that the
singular Landau level structure strongly affects the renormalised gap
in the intermediate coupling regime, which is relevant to experiments
on the fractional quantum Hall effect.   We compare our findings
with measurements of the gap in fractional Hall states with
filling fraction $\nu$ near $\nu=1/2$.
\end{abstract}


\vskip 1cm

]
\narrowtext
\tightenlines

\subsection*{1. Introduction}
One of the most startling discoveries in low-temperature physics in the
last few decades was the identification of a new collective state of
interacting 2-dimensional electrons in semiconducting films or
heterostructures, for temperatures $T<1$K.   
We now know that a whole hierarchy of ``fractional quantum Hall" states
exists, at special fields such that the ``filling fraction"
$\nu$ of electrons in the lowest Landau level is equal to a set of simple 
fractions (the ``fractional hierarchy" of states);
usually these occur for $\nu<1$, but have also been observed for
$\nu >1$.

Theoretical approaches to this new state of matter started with
Laughlin's ``wave-functions" for the ground state and single 
quasi-particles; more recent years have seen several attempts at
field theoretical descriptions of the low-energy physics.
The Laughlin theory \cite{laugh} of the fractional quantum Hall effect 
(FQHE) very successfully describes the ground state and low-energy 
quasiparticles near the filling fractions $\nu_k = 1/(2k+1)$, with
$k= 1,2,3 \ldots$   Nevertheless for at least a decade theorists have been 
searching for an effective field theory for the FQHE, for various reasons,
amoungst which are

(i) The need for a theory which deals not only with filling fractions
$\nu_k$, but for the full experimentally observed hierarchy of states.

(ii)  The need to calculate correlation functions for the low-energy physics
near all the stable filling fractions.

(iii)  The need to deal with situations where the quasi-particle density
is not particularly low, especially in regions where potential fluctuations
make the Laughlin wave-function difficult to apply.

The main difficulty is of course the combination of strong correlations
and Landau level degeneracy - switching on the interactions must lead to a 
non-analytic (in the coupling strength) change in the spectrum.

A bosonic field theory \cite{zhang} of the FQHE introduced a Chern-Simons
field to convert the electrons to bosons, and treated the ensuing field theory
in a mean-field approximation; the FQHE ground states at $\nu = \nu_k$ 
then map onto superfluids \cite{girvin}, while the quasiparticles map
onto vortices.  Unfortunately, despite some success, this theory 
cannot deal with the other stable filling fractions ($\nu \neq \nu_k$).

A more recent fermionic theory, devised by Jain \cite{jain1}, attaches an 
{\em even} number of artificial ``statistical" flux quanta to each electron.
This theory describes FQHE states with filling fraction $\nu = \nu^m_p = 
p/(2mp+1)$,
where $p = \pm1, \pm2, \pm3, \ldots$, and $m = 1,2,3,\ldots$; the number
of statistical flux quanta attached to each electron is then $2m$.
Interestingly, the states generated appear to correspond much better to the
experimental hierarchy; moreover, if we let $|p| \rightarrow \infty$, so 
that $\nu \rightarrow 1/2$, a number of interesting predictions arise.
At $\nu =1/2$ the external magnetic field, $B_{1/2}$,
is such that there are two magnetic
flux quanta for each electron.  Electrons are transformed into 
composite fermions  by attaching two flux quanta to each in
such a way as to cancel on average the field $B_{1/2}$.  
At $\nu = p/(2p+1)$ (away from $\nu = 1/2$) 
 the CF's find themselves 
in a field $\Delta B = B_{\nu}- B_{1/2}= n_e((2p+1)/p-2) = n_e/p$.
Thus we are led to the observation 
by Jain that the FQHE
of electrons is the integer QHE of composite fermions.

Various methods have been used to determine the cyclotron radius of 
the charged quasiparticles away from $\nu = 1/2$ \cite{cyclo}. 
These experiments have confirmed that these quasiparticles do in fact behave
as if they were in a magnetic field $\Delta B$ (so that the cyclotron 
radius diverges as $B \rightarrow B_{1/2}$), which corresponds
precisely to the prediction of CF theory.
In addition to this, the theory of CF's also predicts a gap of size
$e\Delta B/m^{*}$, where $m^{*}$ is the effective mass of the CF's.
There have been numerous direct measurements of the FQHE gap for very small
values of $p$ ($p=1,2$) but for slightly larger values of p an indirect approach
has been used.   The effective mass is found by fitting the Shubnikov-de Haas
(SdH) 
oscillations to the Dingle
formula for non-interacting quasiparticles,
parametrized by a variable effective
mass and by an energy-independent scattering rate.  Du. et. al. have
reported that the effective mass appears to diverge for large values of
$p$ \cite{du}.  However it should be noted that another group has
reported different results \cite{leadl}, up to a smaller value of $p$.
Nevertheless, as $\Delta B \rightarrow 0$, when $\nu \rightarrow 1/2$,
the gap is observed experimentally to go to zero, ie., $m^{*}(\Delta B)$
must diverge more slowly than $1/\Delta B$, if at all.

The apparent success of the Jain theory led to its elaboration into a 
``CF gauge theory", where the statistical gauge flux is again introduced
via a Chern-Simons field \cite{lopez,hlr}.  The consequences of 
this theory have been examined both at $\nu = 1/2$, in the gapless
state, and also away from $\nu = 1/2$, near the filling fractions 
$\nu = \nu^1_p = p/(2p+1)$.  In both cases one must deal with the long-
range dynamic gauge interaction between the CF's. The exact form of this 
gauge propagator depends on the form of the original electron-electron 
interactions.
When the electron-electron interactions are unscreened Coulombic the  
theory of CF's  describes a marginal Fermi liquid at $\nu = 1/2$.
There is a well defined 
Fermi surface but the quasiparticles of the theory cease to be
well defined in the vicinity of the Fermi energy, because of 
a large imaginary self-energy.
This coincides with a logarithmic divergence  in the
effective mass  near the
Fermi surface \cite{hlr}. 

When the interaction between the electrons is  screened, the gauge
interaction between the CF's becomes very severely infra-red divergent.
Despite a large theoretical effort to understand this problem at $\nu = 1/2$,
using various non-perturbative methods \cite{theory}, there is no agreement 
on the final form of the theory for this case. 

Away from $\nu=1/2$ the gauge theory is easier to control, since the CF
gap provides an IR cut-off.  Quite surprisingly, in view of the large
effort devoted to the $\nu = 1/2$ system, few papers have attempted to apply 
the CF gauge theory to the FQHE states.  Kim et. al. \cite{kim1} calculated
both the CF gap and the finite-T compressibility at low T, 
and Stern and Halperin \cite{stern} examined the CF gap and the chemical 
potential. Simon and Halperin \cite{simon} also calculated, using the
RPA, an expression for the current-current correlation function.
Curnoe and Stamp \cite{curnoe} gave a preliminary examination of the
CF spectral weight, and compared this with what one expects for the 
simpler problem of 2-d electrons, moving in Landau levels, interacting with
ordinary phonons.   These investigations were all perturbative; and moreover
used the approximation $|p| \gg 1$, allowing a semi-classical treatment 
of the sums over Landau levels.  All of these papers studied the CF mass
in the FQHE states.

The present paper has two purposes. First, we give a more
detailed discussion of the brief results presented in \cite{curnoe}.
Second, and more important,
we wish to give a more careful treatment of the divergences,
present in $\Sigma(\epsilon)$, which come from the Landau level 
quantization.  As noted in \cite{curnoe}, these divergences have a rather
peculiar structure in the interacting gauge system, which led us 
to look at the corresponding structure in the simpler electron-phonon 
problem as well.   Here we try and give a self-consistent treatment of 
this structure, going beyond perturbation theory.
As we shall see, it is important to do this, since a 
proper account of it leads to important changes in the dependence of the 
FQHE gap in the filling fraction.   We give both approximate and
analytic results; our method is essentially a 
 sum of rainbow graphs,
for CF's moving in Landau levels, and interacting via a singular
gauge field.

Our main results are as follows.  On the purely theoretical side,
we show how one may calculate the forms of both the self-energy
and the quasi-particle spectral function, for the case of both 
electrons interacting with phonons, and composite fermions 
interacting with gauge fluctuations; in both cases these calculations
are done in {\em finite field}, incorporating the singular Landau
level effects.  It turns out that a self-consistent treatment of
these effects is not entirely obvious, and so we show how this may 
be done.

We then apply our technique to the calculation of the experimentally
measurable gap, again for the electron-phonon problem and the
composite fermion problem.  Our results show that if one does not 
handle the singular Landau level structure self-consistently,
the gap in the experimentally relevant regime is badly
overestimated in the case of the FQHE; we give results both for 
CF's and the electron-phonon problem.

The paper is organised as follows.  In section 2 we discuss the self-energy 
and spectral function of Landau level electrons coupled to phonons.  This
problem is studied to throw light on the more complex CF problem.  It is 
also of purely academic interest - as far as we are aware, despite
the enormous literature on the electron-phonon system, no previous study 
has been made of the 1-particle propagator
 of electrons in Landau levels, coupled 
to phonons which deals with the singular structure introduced by
Landau level quantization \cite{note}.  The calculations are carried out for both
Debye and Einstein phonons. 

In section 3 we describe in detail our lowest-order perturbation
calculation of $\Sigma(\epsilon)$ and $A(\epsilon)$ for CF's in Landau
levels, interacting via a gauge field.  In both the electron-phonon 
and CF gauge cases, one may distinguish a strong- and weak-coupling 
regime.  In the weak-coupling regime the Landau level excitation gap is 
hardly affected, whereas it is strongly renormalised in the strong-coupling
regime.

The non-trivial structure of $A(\epsilon)$, the quasiparticle spectral
function, leads us in section 4 to introduce an iterative self-consistent
calculation of $\Sigma(\epsilon)$, which takes systematic account of the
coherent parts of $\Sigma(\epsilon)$.  This then allows us to give an
exact rainbow graph summation of these coherent parts, and we then use this 
compute the CF gap over a wide range of coupling constants and filling
fractions.

\subsection*{2. Phonon Interactions}
\begin{figure}[h]
\epsfysize=1.0in
\epsfbox[42 195 558 436]{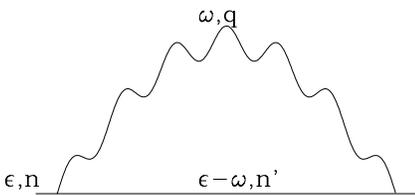}
\caption{Feynman diagram representing the self-energy. $n$ and $n^{'}$ are
Landau level indices.  The straight line represents the fermion
(either electron or composite fermion) and the wavy line
represents the boson (either a phonon or a gauge fluctuation).
\label{fi:selfe}}
\end{figure}

We begin by examining the self-energy $\Sigma_n(\epsilon)$ and the spectral
function $A_n(\epsilon)$ for electrons coupled to phonons, whilst moving 
on Landau level $n$ (see Fig. \ref{fi:selfe}).  The structure introduced
by the combination of interactions and Landau level degeneracy turns out 
to be rather interesting.  

In {\em zero} field, the 
 spectral functions of the coupled electron-phonon system were first
studied by Engelsberg and Schreiffer \cite{engel}. 
 Using  optical and Debye phonon
spectra they calculated the real and imaginary parts of the self-energy
using conserving approximations (obeying the Ward identities).
They found that the spectral function,
\begin{eqnarray} 
A(\epsilon,{\bf p})& =& \frac{1}{\pi}\mbox{Im}G(\epsilon,{\bf p}) 
\nonumber \\
          & =&  \frac{1}{\pi}\frac{\Sigma^{''}(\epsilon,{\bf p})}
		 {(\Sigma^{'}(\epsilon,{\bf p}) 
                  +\epsilon-\xi_{\bf p})^2 + \Sigma^{''}(\epsilon,{\bf p})^2},
		  \label{eq:ImagG}
\end{eqnarray}
yielded structures that they could identify as well defined qp's
for two cases: when {\bf p} is close to the Fermi surface, and when it 
is very large.  In the intermediate regime there exists 
only an incoherent ``smeared" structure.

In our analysis we first 
consider a two dimensional system of electrons in a
small magnetic field $B$
which interacts with phonons having a Debye spectrum.
In this case we will use a the phonon propagator calculated at $B=0$,
\begin{equation}
D(q,\omega) = \frac{q c_s}{\omega^2 - c_s^2q^2}.
\label{eq:debye}
\end{equation} 
which is assumed to have a cutoff at the Debye frequency $\omega_D$
($c_s$ is the speed of sound).
Within this same semiclassical limit we adopt a well known
approximation 
for the overlap matrix element, between plane waves and Landau level states 
(see the appendix for more details):
\begin{equation}
         |\Lambda(n,n^{'},q)|^2  =    \frac{1}{q\pi}
        \left(\frac{eB}{2n}\right)^{1/2} \label{eq:matrix}
\end{equation}
where $n$ and $n'$ are the Landau level indices as shown in Fig.
\ref{fi:selfe}, and  $n\approx p$ is the number of filled Landau levels.
The self-energy expression is given in perturbation theory by 
\begin{eqnarray}
    \Sigma_{n}(\epsilon)&  =&
                   \int \frac{d^2q}{(2\pi)^2} \int_{0}^{\infty}
                    \frac{d\omega}{\pi}
 {\rm Im}U(q,\omega) \nonumber \\
             & &       \sum_{n^{'}=0}^{\infty}
     |\Lambda(n,n^{'},q)|^{2}\int_{-\infty}^{\infty}\frac{d\epsilon^{'}}{\pi}
                 \mbox{Im}G^{0}_{n^{'}}(\epsilon^{'}) \nonumber \\
        & \times &     \left(\frac{1+n_{B}(\omega)-n_{f}(\epsilon^{'})}
             {\epsilon-\epsilon^{'}-\omega+i\delta}
      + \frac{n_{B}(\omega)+n_{f}(\epsilon^{'})}{\epsilon -\epsilon^{'}+\omega+
                       i\delta} \right), \label{eq:selfe}
\end{eqnarray}
where $G^0$ is the bare electron Green function, ie.,
\begin{equation}
 \mbox{Im }  G^0_n(\epsilon) = \pi\delta(\epsilon+(p-1/2-n)\omega_c).
\end{equation}
and $U(q,\omega)$ depends on the electron-phonon coupling (see below).
The electronic frequency $\epsilon$ is measured from the Fermi
energy, $\mu = p \omega_c$, and the energy of the highest filled
Landau level is $(p-1/2)\omega_c$; since we are doing a quasi-classical
calculation, we assume that the Fermi energy is halfway between Landau
levels \cite{hlr,kim1}$^-$\cite{curnoe}.

In the rest of this section we first calculate the self-energy
and the renormalised cyclotron gap energy for  the case of a 
deformation coupling to Debye  phonons; we then do the same for
the coupling to optical Einstein phonons.

{\bf (a) Debye Phonons:}  We start with Debye phonons having 
velocity $c_s$.  The function $U(q,\omega)$ in (\ref{eq:selfe})
is just the effective interaction between electrons; in the Debye
model it is given by
\begin{equation}
     U(q,\omega) = \frac{q \Xi_D^2 \hbar}{2 c_s \rho a}D(q,\omega).
\end{equation}
where $D(q,\omega)$ is the phonon Green function (\ref{eq:debye}).
The electron-phonon
interaction is parametrised by the deformation potential, $\Xi_D$, and 
$\rho$ is the ion mass 
density of the material (see the appendix for more details).

Evaluating the integrals in eq. (\ref{eq:selfe}) yields, at temperature 
$T=0$:
\begin{eqnarray}
\Sigma_{n}^{'}({\epsilon}) & =  &
              \frac{K_D}{\pi}     \left(
                \sum_{n^{'}=p}^{\infty}\left(-\omega_D+
		 ((n^{'}-p+1/2)\omega_c-{\epsilon})\right. \right.\nonumber \\
& & \left. \left. \log\left|
    \frac{\omega_{D}+(n^{'}-p+1/2)\omega_c-{\epsilon}}
         {(n^{'}-p+1/2)\omega_c-{\epsilon}}
		  \right|\right)\right. \nonumber \\ 
 &+& \left. \sum_{n^{'}=0}^{p-1} 
      \left(\omega_D+((n^{'}-p+1/2)\omega_c-\epsilon)\right.\right. \nonumber \\
& & \left. \left. \log\left|
            \frac{(n^{'}-p+1/2)\omega_c-\epsilon-\omega_D}
           {(n^{'}-p+1/2)\omega_c-\epsilon}\right|
	    \right)\right)
\end{eqnarray}
\begin{eqnarray}
\Sigma_{n}^{''}(\epsilon)&  = &
     K_D\left(\sum_{n^{'} = p}^{\infty} (\omega_D-\epsilon \right. 
\nonumber \\
   +  ((n^{'}&-&p+1/2)\omega_c-\epsilon)
	  \theta((n^{'}-p+1/2)\omega_c))  \nonumber \\
& &     + \sum_{n^{'} = 0}^{p-1}
    (\omega_D+\epsilon  \nonumber \\
   +       ((-n^{'}&+&p-1/2)\omega_c+\epsilon)
     \left.     \theta((-n^{'}+p-1/2)\omega_c)) \right),
\end{eqnarray}
The right hand side
is independent of $n$, and so we will omit the subscript on $\Sigma(\epsilon)$
henceforth. $\Sigma(\epsilon)$ is in units of $\omega_c$ and
$K_D$ is a dimensionless
constant,
\begin{equation}
K_D = \frac{\Xi^2_D m \omega_c}{4 \pi c_s^3 a\rho}
     \left(\frac{1}{4\pi n_e}\right)^{1/2} \label{eq:KD}
\end{equation}

\begin{figure}[h]
\epsfysize=3.2in
\epsfbox[0 154 558 683]{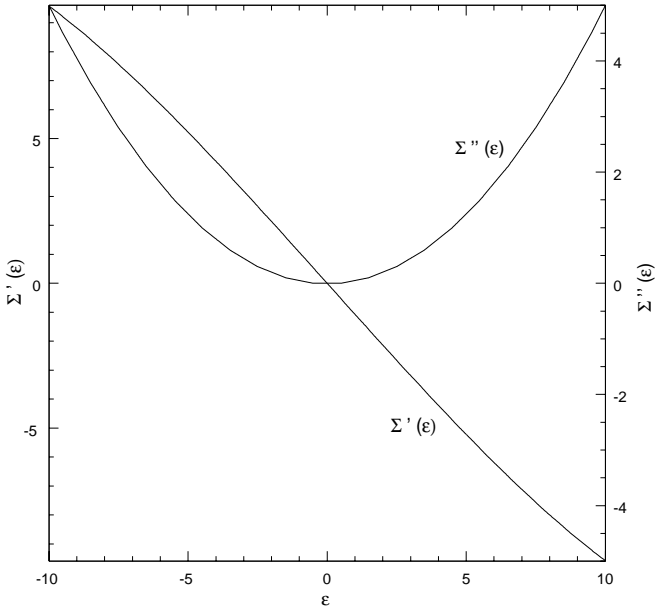}
\caption{The real and imaginary parts of the self-energy for electrons
interacting with phonons with a Debye spectrum ($\omega_D = 20\omega_c$)
in a small magnetic field.    All energies are in units of $\omega_c$;
the coupling is $K_D=1$.
 \label{fi:selfephonon}}
\end{figure}

It is useful to rewrite these equations shifting the sum to start at $-p$
and then enforcing particle-hole symmetry by truncating the sum at $p-1$, the
number of filled Landau levels. (In reality the problem is not 
particle-hole symmetric, but we are only interested in energies
within $\omega_c$ of the Fermi surface, where particle-hole
symmetry is almost exact. In the case of composite fermions, this 
will correspond to having an upper cutoff at the true cyclotron frequency).
This gives:
\begin{eqnarray}
& &\Sigma^{'}(\epsilon)  =  \frac{K_D}{\pi} \sum_{m=0}^{p-1}
                       \nonumber \\
& & \left(       ((m +1/2)\omega_c -\epsilon)
          \log \left|\frac{(m+1/2)\omega_c-\epsilon+\omega_D}
              {(m+1/2)\omega_c-\epsilon}\right|
      \right.  \nonumber \\
& & \left.      -((m+1/2)\omega_c+\epsilon)
        \log\left|\frac{(m+1/2)\omega_c+\epsilon+\omega_D}
                         {(m+1/2)\omega_c+\epsilon}\right|\right)
\label{eq:rselfephD}
\end{eqnarray}
\begin{eqnarray}
\Sigma^{''}(\epsilon)& =& K_D\sum_{m=0}^{p-1} 
(((m+1/2)\omega_c-\epsilon)  \nonumber \\
    \times   \theta((m&+&1/2)\omega_c-\epsilon +\omega_D)
         \theta((-m-1/2)\omega_c+\epsilon) 
           \nonumber \\
&    + & ((m+1/2)\omega_c+\epsilon)
  \nonumber \\
\times  \theta((m&+&1/2)\omega_c+\epsilon+\omega_D)
       \theta((-m-1/2)\omega_c-\epsilon))
\label{eq:iselfephD}
\end{eqnarray} 
These functions are shown together in Fig. \ref{fi:selfephonon}.  The 
structure caused by the Landau levels is very weak; it can just be seen in
$\Sigma^{''}(\epsilon)$ (if $\omega_c/\omega_D$ is larger, it is much more 
obvious).   $\Sigma^{''}(\epsilon)$ is simply a sum of ramp functions, 
coming from each Landau level; as $\omega_c \rightarrow 0$, 
$\Sigma^{''}(\epsilon)$ becomes parabolic for $\epsilon \ll \omega_D$.
The logarithmic singularities in $\Sigma^{'}(\epsilon)$ are quite invisible
in Fig. \ref{fi:selfephonon}.

\begin{figure}[h]
\epsfysize=3.3in
\epsfbox[21 145 585 690]{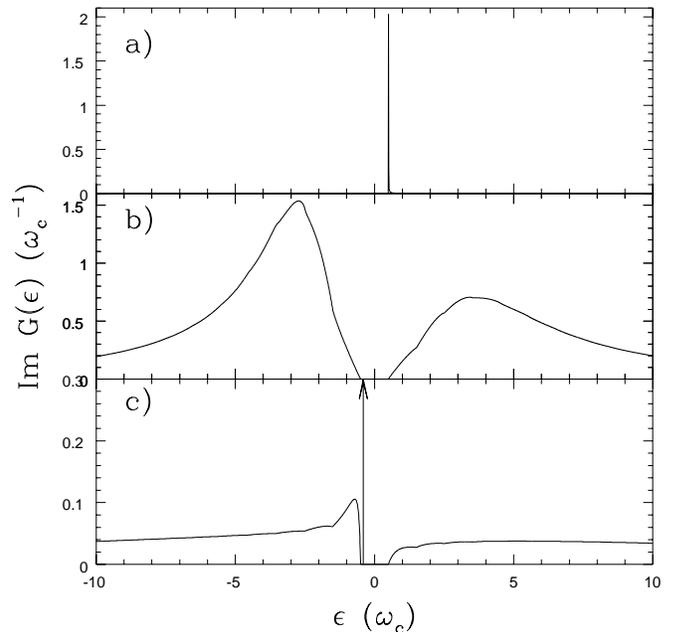}
\caption{The imaginary part of the Green function
 of electrons interacting with Debye phonons
for $K_D=.001,.1$ and .25  with $n=0$ (see the text for definition of the
coupling $K_D$).
\label{fi:specfl}}
\end{figure}

We wish to find the poles of $G_n(\epsilon)$.
Generally there are 0, 1, 2 or 3 roots to the equation 
\begin{equation}
\Sigma^{'}(\epsilon)+\epsilon = (n+1/2)\omega_c,  \label{eq:solve}
\end{equation} 
which correspond to peaks in the spectral function, Im$G_{n}(\epsilon)$.
In this expression both $n$ and $\epsilon$ are measured with respect to 
the Fermi surface.  We are only concerned with the low energy
behaviour which corresponds to small $n$ and $\epsilon$.
When $n$ is small and
$K_D$ is small there is one solution occurring near 
$\epsilon = (n+1/2)\omega_c$,
which appears as a well defined quasiparticle peak (a $\delta$-function for 
$n=0$).
When $K_D$ is
large there are three solutions.  Two 
occur well beyond the Debye frequency (outside of the low energy 
regime) and the third occurs close to
$\epsilon = 0$ yielding a $\delta$-function.   Intermediate values of $K_D$
do not in
 general yield the very narrow peaks characteristic of a dressed particle, 
since if solutions to (\ref{eq:solve}) do occur they occur at finite $\epsilon$
where the imaginary part of $\Sigma$ is also finite.
There may also be features  arising from incoherent 
contributions.  
Some of
these are shown in Fig. \ref{fi:specfl}.  We remark  that the 
magnetic field causes slight discontinuities in the slope of 
the spectral functions at the Landau level energies.

\begin{figure}[h]
\epsfysize=1.9in
\epsfbox[40 389 562 686]{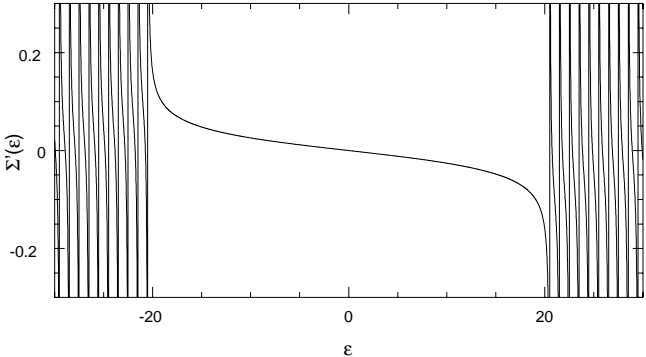}
\caption{Real part of the self-energy of electrons interacing
with phonons with an Einstein spectrum ($\omega_E = 20 \omega_c$).  Both
axes are in units of $\omega_c$; the coupling is $K_E=1$.
\label{fi:selfeinstein}}
\end{figure}

{\bf (b) Einstein Phonons:}
We now consider the interaction with the more singular Einstein phonon
spectrum, with phonon propagator $D(q,\omega) = 
\omega_E/(\omega^2 - \omega_E^2)$; $\omega_E$ is the ``optical" phonon
frequency.  In {\em zero field}, this was first
considered by Engelsberg and Schreiffer
who found the self-energy to be
\begin{eqnarray}
   \Sigma^{'}(\epsilon) & = & -\frac{K_E\omega_c}{\pi}
           \log\left|\frac{\epsilon+E_F+\omega_E}{\epsilon +E_F
                    -\omega_E} \right|   \\
 \Sigma^{''}(\epsilon) & = & K_E \omega_c \hspace{.5in} \mbox{for} |\epsilon| > \omega_E \nonumber\\
                        & = & 0 \hspace{.7in} \mbox{for} |\epsilon| <\omega_E
\end{eqnarray}
With $\Sigma$ expressed in units of $\omega_c$,
$K_E$ is the dimensionless coupling:
\begin{equation}
K_E \approx \frac{g^2 a (4\pi n_e)^{1/2}}{4 E_F \omega_c}.
\end{equation}
The effect of the magnetic field is to modify these equations to
\begin{eqnarray}
 \Sigma^{'}(\epsilon) & = & -\frac{K_E \omega_c^2}{\pi}\sum_{m = 0}^{p-1}\left(
        \frac{1}{(m+1/2)\omega_c+\omega_E-\epsilon} \right. \nonumber \\
  & &\left. 
     \hspace{.8in}   - \frac{1}{(m+1/2)\omega_c+\omega_E+\epsilon}
        \right)  \label{eq:rselfephE}
\end{eqnarray}
\begin{eqnarray}
  \Sigma^{''}(\epsilon) & = & K_E \omega_c^2 \sum_{m = 0}^{p-1}
      (\delta((m+1/2)\omega_c+\omega_E-\epsilon) 
\nonumber \\
& &  \hspace{.6in}  +\delta((m+1/2)\omega_c+\omega_E+\epsilon))
       \label{eq:iselfephE}
\end{eqnarray}
$\Sigma^{'}(\epsilon)$ is shown in Fig. \ref{fi:selfeinstein}.
There is no virtually no effect of a small magnetic field on the
self energies for $|\epsilon|<\omega_E$; beyond this region  the
smooth functions are replaced by functions with divergences at each
Landau level energy. All poles in the spectral function appear as $\delta$-functions
since the imaginary part of the self-energy
 vanishes everywhere except at discrete
points, but the weights of the poles vary.
The new electron
spectrum is determined by summing over all $n$. Since the $n$ are
discrete a renormalised gap can be defined as the energy
difference between the highest occupied and lowest unoccupied
states, ie., by the difference between the positions of the
$\delta$-function peaks of $G_0(\epsilon)$ and $G_{-1}(\epsilon)$.
 
The amount of renormalization of the gap  depends
on the size of the coefficient of the self energy, $K_E$.  When $K_E$ is
small the poles lie near $(n+1/2)\omega_c$ 
for $n\omega_c<\omega_E$, as in the Debye case.  The
renormalized gap is approximately  $\omega_c$, only slightly increased due
to the interactions.  As $K_E$ is increased the gap increases further.
The gap may be determined by approximating
\begin{equation}
\Sigma^{'}(\epsilon) = \frac{\partial \Sigma^{'}(0)}{\partial\epsilon} \epsilon
\label{eq:approx}
\end{equation}
and then solving (\ref{eq:solve}) for $n=-1$ and $n=0$:
\begin{eqnarray}
\frac{\partial \Sigma^{'}(0)}{\partial\epsilon} &= &
\frac{-K_E}{\pi}\sum_{m=0}^{p-1}\frac{2}{(m+1/2)\omega_c+\omega_E}\\
& \approx & -\frac{2K_E}{\pi}\log\left(
     \frac{(p+1/2)\omega_c+\omega_E}{\omega_c/2+\omega_E}\right)
\end{eqnarray}
where the sum over $m$ has been replaced by an integral.   Then the
solution to (\ref{eq:solve}) is
\begin{equation}
\epsilon_{\pm}\left(1-\frac{2K_E}{\pi}\log\left(\frac{(p+1/2)\omega_c+\omega_E}
      {\omega_c/2+\omega_E}
\right)\right) = \pm\frac{\omega_c}{2}
\end{equation}
and the gap is
\begin{eqnarray}
\omega_c^{*}& =  & |\epsilon_{+} - \epsilon_{-}| \nonumber \\
             &  = & \omega_c\left|1-\frac{2K_E}{\pi}\log\left(
               \frac{(p+1/2)\omega_c+\omega_E}
      {\omega_c/2+\omega_E}\right)\right|^{-1}
               \label{eq:gapE}
\end{eqnarray}
This procedure is also applicable to the Debye case
when there are
well defined quasiparticles for $n = -1,0$.   In this
case the renormalized gap is
\begin{eqnarray}
\omega_c^{*}&  = & \omega_c\left|1
-\frac{K_D}{\pi}\left(\log(2\omega_D+\omega_c) \right. \right.\nonumber \\
&  &-  (2p+1)
  \log\left.\left.\left(\frac{(p+1/2)\omega_c+\omega_D}{(p+1/2)\omega_c}\right)
    \right)\right|^{-1} \label{eq:gapD}
\end{eqnarray}
We may  draw some general conclusions
about the size of the gap for different coupling strengths using the
forms above.  Considering first the Einstein case,
for small values of the couplings the gap is
renormalized to a larger value.  This continues to be the case as
the coupling is increased.   The divergence in the gap seen in
equation (\ref{eq:gapE}) is avoided because the gap is always
bounded by $2\omega_E$.  This is due to the presence of the
logarithmic peaks
in the real part of
$\Sigma(\epsilon)$
which ensures that there will always be a solution to (\ref{eq:solve})
for $|\epsilon|<\omega_E$.  For large couplings the gap is reduced.
In the Debye case, for small $K_D$ there are well defined quasi-particle
peaks outside the gap but they are not strictly $\delta$-functions
thanks to small but non-zero Im$\Sigma(\epsilon)$.  The finite width
maintains the size of the gap to be $\omega_c$, as shown in Fig.
\ref{fi:specfl}.  For larger values of $K_D$ the divergence of the gap in
(\ref{eq:gapD}) is avoided for the same reason, as shown in Fig.
\ref{fi:specfl}b.  For large couplings the gap is reduced because of
q.p. poles that lie within $\omega_c$, as shown in Fig. \ref{fi:specfl}c.

As can be seen from Figs. \ref{fi:selfephonon} and \ref{fi:specfl},
the effect of Landau quantization on the quasi-particle properties,
for realistic values of $\omega_D, \omega_E$ and $\omega_c$,
is very small, at least at low energies   (it is however worth
noting that once $\epsilon>\omega_E$ in the case of Einstein phonons,
a rather obvious singular structure appears in the self-energy,
see Fig. \ref{fi:selfeinstein}).   Nevertheless the calculation
of these effects may be useful in systems for which $\omega_c$ can
be made as large as $\omega_D$ or $\omega_E$.

Our main reason for looking at the electron-phonon problem is that
it is a well-understood example of a Fermi liquid, which nevertheless
acquires a non-trivial structure when Landau quantization is introduced.
  We now turn
to composite fermions, which are not so simple - the results 
just derived will help us to understand the CF results.

\subsection*{3. Composite Fermions and Gauge Fluctuations}
 
We now examine
the spectral functions of composite fermions
near $\nu = 1/2$ in the same manner
as for electrons.  We assume that the Lagrangian has the
usual form for composite fermions interacting with a gauge field
\cite{hlr}$^-$\cite{curnoe}:
\begin{eqnarray}
{\cal L} &= &\int d^2x \psi^{\dagger}(x)(-i\partial_0 + a_0)\psi(x)
\nonumber \\
& & +\frac{1}{2m}\psi^{\dagger}(x)(-i\partial_i+a_i)^2\psi(x)
             +a_{\mu}D^{-1}_{\mu\nu}a_{\nu}
\end{eqnarray}
where in this case $\nabla \times a = \Delta B$.
The most singular component of the
interaction with the gauge propagator, calculated at
$\Delta B = 0$, is \cite{hlr}
\begin{eqnarray}
     U(q,\omega) & = & - \left| \frac{\vec{k}_f\times\hat{q}}{m}\right|^2
                             D_{11}(q,\omega)  \\
  D_{11}(q,\omega)& = &\frac{q}{\bar{\chi} q^s - i \gamma\omega} 
   \label{eq:gauge}
\end{eqnarray}
This is the transverse component, calculated within the Coulomb gauge.
The exponent $s$ can have values between 2 and 3.  $s=2$  when
there are unscreened 
Coulombic electron-electron interactions and $s>2$ corresponds
to screening those interactions (typically by having the 2-d doped 
semi-conductor near to a conducting plate);  the case $s=3$ arises when 
the interaction is effectively zero-range
(of strength $v$, and range less than the magnetic length $l$). 
The constants are $\gamma = \frac{2n_e}{k_f}$
and $\bar{\chi} = \frac{e^2}{8\pi\tilde{\epsilon}}$ for $s=2$ and
$\bar{\chi} = \frac{1}{24\pi m} + \frac{v}{(4\pi)^2}$ for $s=3$ in this 
random phase approximation \cite{hlr} (we denote the dielectric constant
by $\tilde{\epsilon}$).    In what follows we first 
discuss the self-energy, and then the energy gap for CF's, after it has
been renormalised by gauge fluctuations (in the absence of gauge
fluctuations, the gap is just $\Delta \omega_c = \Delta B/m$, as
discussed in the Introduction).

{\bf (a)Composite Fermion Self-Energy:}  The results for the CF
self-energy were given, without derivation, in a previous note
by us \cite{curnoe}; here we briefly recall these (some
details of the derivation appear in the Appendix).

\begin{figure}[h]
\epsfysize=3.3in
\epsfbox[20 145 567 686]{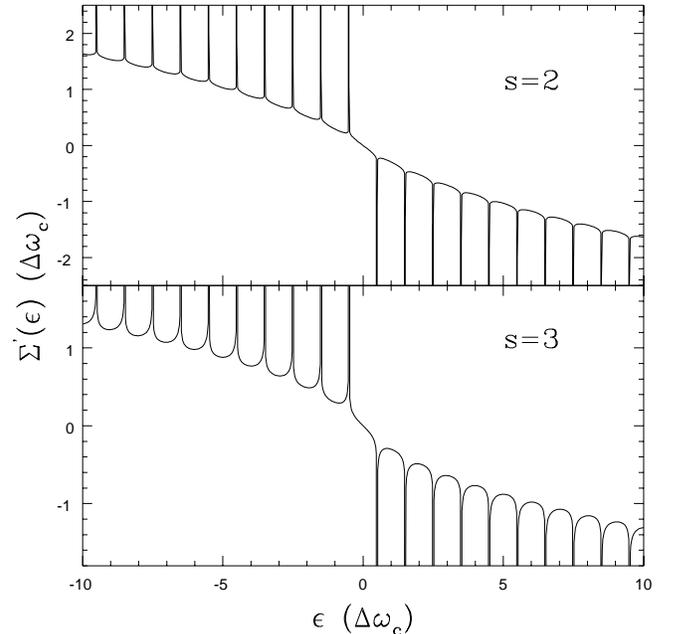}
\caption{Real part of the self-energy of composite fermions. $s=2$
corresponds to Coulombic electron-electron interactions; $s=3$
corresponds to short-ranged screened interactions (see text).  The energies
are in units of $\Delta\omega_c$.
\label{fi:selfecf}}
\end{figure}

\begin{figure}[h]
\epsfysize=3.3in
\epsfbox[20 145 575 690]{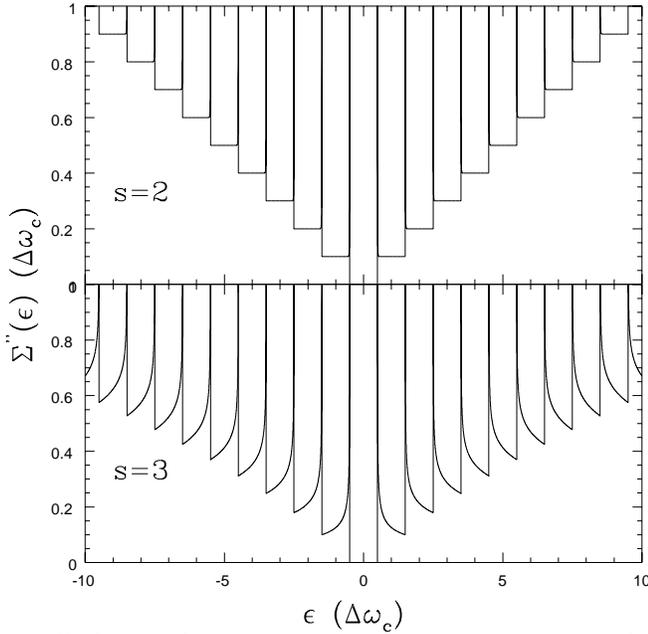}
\caption{Imaginary part of the self-energy  of composite fermions.
$s=2$
corresponds to Coulombic electron-electron interactions; $s=3$
corresponds to short-ranged interactions and the energies are in
units of $\Delta\omega_c$, as before.
\label{fi:selfecf2}}
\end{figure}

We begin by writing down the real and imaginary parts of 
$\Sigma_{CF}(\epsilon)$, at $T=0$ (where they can be written
in analytic form).  For 
$s=2$ one has:
\begin{eqnarray}
\Sigma^{'}(\epsilon) & = &
   \frac{K_2 \Delta \omega_c}{\pi}
 \sum_{m=0}^{p-1}
  \log\left|\frac{(m+1/2)\Delta\omega_c-\epsilon}
          {(m+1/2)\Delta\omega_c+\epsilon}\right| 
      \label{eq:rselfecf2} \\
  \Sigma^{''}(\epsilon) & =&
      K_2 \Delta\omega_c\sum_{m=0}^{p-1}
      (\theta((-m-1/2)\Delta\omega_c-\epsilon) \nonumber \\ +
& &       \theta((-m-1/2)\Delta\omega_c+\epsilon))  
   \label{eq:iselfecf2} \\
K_2 & = & \frac{E_F\tilde{\epsilon}}{e^2\sqrt{4\pi n_e}} = 
    \frac{\tilde{\epsilon}\sqrt{4 \pi n_e}}{2 m e^2} \label{eq:k2}
\end{eqnarray}
whilst for  $3\geq s>2$:
\begin{eqnarray}
 \Sigma^{'}(\epsilon)&  = & {\rm sgn}(\epsilon)\cot\left(\frac
      {2\pi}{s}\right)
      \Sigma^{''}(\epsilon) \nonumber \\
  &   + & \csc\left(\frac{2\pi}{s}\right)K_s\sum_{m=0}^{p-1}
    \nonumber \\
& &(-((m+1/2)\Delta\omega_c-\epsilon)^{-\alpha}
      \theta((m+1/2)\Delta\omega_c-\epsilon) \nonumber \\
    & &  + (m+1/2+\epsilon)^{-\alpha}\theta(m+1/2+\epsilon))
     \label {eq:rselfecfs}
\end{eqnarray}
\begin{eqnarray}
& &     \Sigma^{''}(\epsilon)  = 
         K_s \sum_{m=0}^{p-1}( \nonumber \\
& &     ((-m-1/2)\Delta\omega_c+\epsilon)^{-\alpha}
      \theta((-m-1/2)\Delta\omega_c+\epsilon)\nonumber \\
& &  + ((-m-1/2)\Delta\omega_c-\epsilon)^{-\alpha}
       \theta((-m-1/2)\omega_c-\epsilon))
     \label{eq:iselfecfs}  \\
& &K_s  =   \frac{E_F}{4s\pi\gamma^{\alpha}\bar{\chi}^{2/s}\sqrt{4\pi n_e}}
            \csc\left(\frac{\pi}{s}\right) \label{eq:ks}
\end{eqnarray}
where $\alpha = (s-2)/s$ is positive.   Plots of $\Sigma^{'}(\epsilon)$ and
$\Sigma^{''}(\epsilon)$
for $s=2,3$  are shown in Figs. \ref{fi:selfecf} and \ref{fi:selfecf2}. 

A rather extraordinary feature of the CF propagator appears once we 
closely examine $\Sigma^{''}(\epsilon)$ and $\Sigma^{'}(\epsilon)$ 
around the divergences - we notice that $\Sigma^{'}(\epsilon)$ shows
only positive divergence on {\em both sides} of each Landau level. 
The paradox is that typically one would expect $\Sigma(\epsilon)$
to have the form 
\begin{equation}
\Sigma(\epsilon) \sim \sum_r \frac{|V_r|^2}{\epsilon - E_r +i\delta}
\label{eq:expect}
\end{equation}
in simple perturbation theory if our starting fermion spectrum is
composed of discrete (albeit degenerate) levels at energies $E_r$, and
$V_r$ is some perturbation.  This leads to divergences in 
$|\Sigma^{'}(\epsilon)|$, but we note that these divergences {\em change sign}
each each time one crosses the energies $E_r$. 

\begin{figure}[h]
\epsfysize=3.4in
\epsfbox[25 145 572 690]{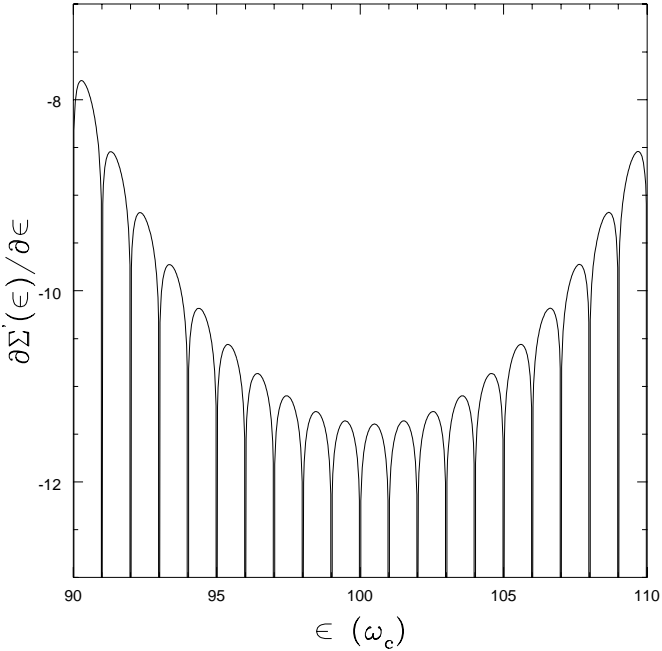}
\caption{The derivative with respect to energy of the real
 part of the self-energy of electrons interacting with Debye phonons is a finite
magnetic field  ($\omega_D = 20 \omega_c$). The energy is in units of
$\omega_c$, and $K_D=1$.
\label{fi:derivselfe}}
\end{figure}

This sign change occurs for $\Sigma^{'}(\epsilon)$ in the electron-phonon
problem; this is most clearly seen by plotting 
$\partial\Sigma^{'}(\epsilon)/\partial \epsilon$ for this problem, where 
one sees a series of positive divergences around each Landau level, with no
sign change, as shown in Fig. \ref{fi:derivselfe}.
This is exactly what we would expect from (\ref{eq:expect}),
since it yields
\begin{equation}
\frac{\partial\Sigma^{'}(\epsilon)}{\partial \epsilon} 
   \sim -\sum_r \frac{|V_r|^2}{(\epsilon - E_r +i\delta)^2}
\end{equation}

The behaviour of $\Sigma_{CF}(\epsilon)$ is thus exactly the opposite of
what one expects.   The explanation of this paradox is that although 
$\Sigma_{CF}^{''}(\epsilon)$ is strongly peaked (indeed divergent)
around each Landau level, and does
not change sign as one crosses a Landau level, nevertheless the peak has 
such a peculiar shape that its  Hilbert transform $\Sigma_{CF}^{'}(\epsilon)$
also does not change sign as one crosses a Landau level.  Each Landau
level $r$ contributes  a term  $\sim (\epsilon - r)^{-\alpha}\theta(\epsilon-r)$
to $\Sigma_{CF}^{''}(\epsilon)$, with long tails for $(\epsilon - r) \gg 1$.
The Hilbert transform of such a function does {\em not} change sign as 
$\epsilon$ crosses $r$, unless $\alpha >1/2$; however in the CF gauge 
theory, $0 \leq \alpha \leq 1/3$.

This explains the paradoxical form of $\partial \Sigma/\partial\epsilon$
for composite fermions - it comes from the very long ``tails" which
extend out from each Landau level.   These have no counterpart
in the self-energy of a Fermi liquid, such as the electron-phonon
problem discussed above.

{\bf (b) Composite Fermion Energy Gap:}   We now evaluate the
renormalised energy gap $\Delta \omega_c^{*}$, using our
perturbative result for $\Sigma^{'}(\epsilon)$ above.
In this 
problem the bare CF's have a gap $\Delta\omega_c = \Delta B/m$
separating the occupied and unoccupied states.   We look again to 
equation (\ref{eq:solve}) to find the location of the poles after
the self energy has been included in the CF propagator.  For each value of $n$
there can be many solutions to (\ref{eq:solve}); in particular we note 
that for each value of $n$ there is exactly one pole within the
gap $|\epsilon|< \frac{\Delta\omega_c}{2}$.  Since the imaginary part is zero in this 
region these poles appear as $\delta$-functions in the spectral function.
Using exactly the same approximations as in the electron-phonon calculation,
(\ref{eq:approx}) to (\ref{eq:gapE}),
we find that for $s=2$ the renormalised gap is
\begin{equation}
\Delta\omega_c^{*} \approx
       \Delta\omega_c \left|1-\frac{2K_2}{\pi}\log(2p+1)\right|
^{-1} \label{eq:gapcf}
\end{equation}
and for $s>2$
\begin{equation}
\Delta\omega_c^{*} \approx 
         \Delta\omega_c \left|1-\frac{4K_s}{\sqrt{3}}p^{-\alpha}
      \right|^{-1}.
\label{eq:gapcf2}
\end{equation}
Similar results have been obtained by Stern and Halperin \cite{stern} and
by Kim et. al. \cite{kim1}.  

However the use of equation (\ref{eq:approx}) to get a solution for
(\ref{eq:solve}) is only valid if $\Sigma^{'}(\epsilon)+\epsilon$ varies slowly
over $\epsilon$ such that  $|\Sigma^{'}(\epsilon)+\epsilon|<
\frac{\Delta \omega_c}{2}$, 
and thus (\ref{eq:approx}) is a poor
approximation when $K_2$ is in an intermediate range ($\approx 0.5 $).
Moreover, the divergence in (\ref{eq:gapcf}) at $K_2 = \pi/(2\log(2p+1))$
is unphysical; instead we expect the actual gap $\Delta \omega_c^{*}$ to
decrease monotonically 
with $K_2$.
In fact, we estimate that in the 
actual experiments so far done \cite{du}, $K_2 \approx 0.8$,
which places it right in the intermediate range, so it is necessary to go 
beyond the estimates in (\ref{eq:gapcf}) and (\ref{eq:gapcf2}).

\begin{figure}[h]
\epsfysize=3.4in
\epsfbox[20 139 562 688]{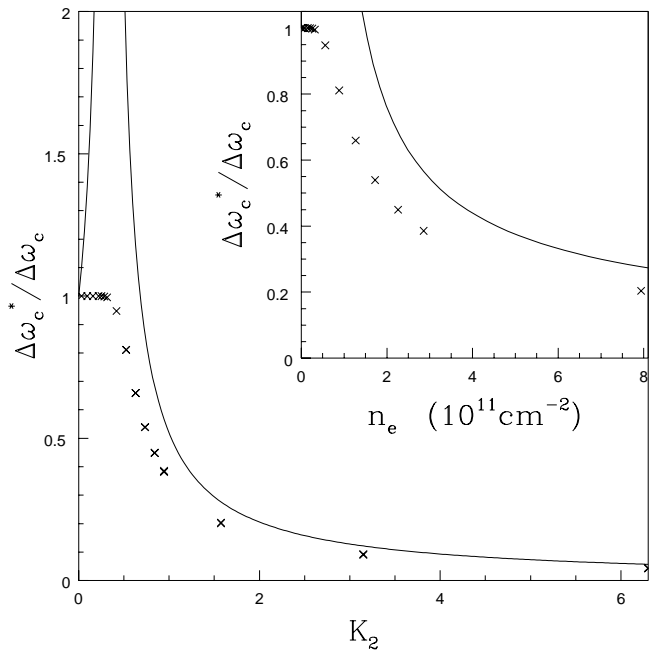}
\caption{First order perturbation results for
the effective gap, $\Delta\omega_c^{*}$, as a function of coupling,
 $K_2$, for
$p=50$.
The solid curve is  $|1-\frac{2K_2}{\pi}\log(2p+1)|^{-1}$
and  points are
numerical solutions to eq. (12).
The inset shows the same data plotted as a
function of $n_e$, which is related to $K_2$ via eq. (29).
\label{fi:gap1cf}}
\end{figure}

\begin{figure}[h]
\epsfysize=3.2in
\epsfbox[7 145 580 686]{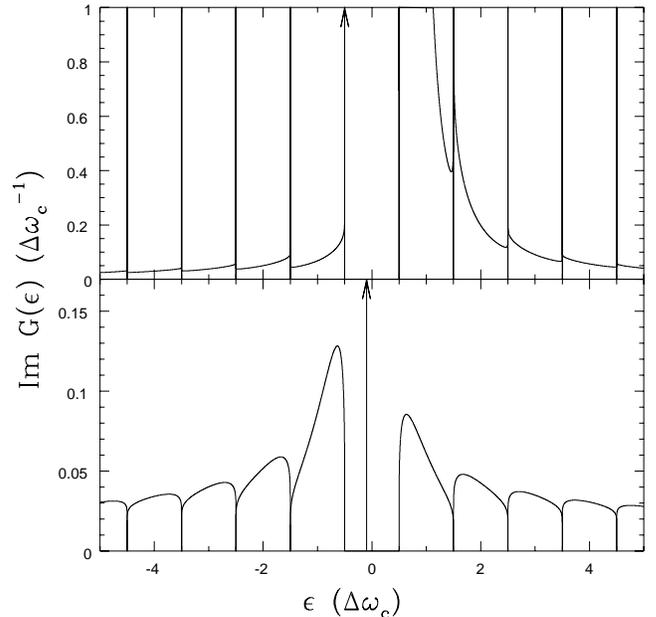}
\caption{Imaginary part of the Green function
 of composite fermions in a finite magnetic
field with $n=0$ and $p=50$.  The upper figure has a coefficient $K_2 = 0.31$
and the lower one has $K_2 = 6.3$. The energy is in units of $\Delta\omega_c$.
\label{fi:speccf}}
\end{figure}

We do this first by solving (\ref{eq:solve})
numerically; as an example we have done this for $p=50$, as shown in Fig.
\ref{fi:gap1cf}. 
The numerical results show that for small values of the
coefficient the gap is reduced by a very small amount.  For larger values
of $K_2$ the gap decreases rapidly.  Turning our attention to
the spectral function shown in Fig. \ref{fi:speccf} we see that
there is a simple physical interpretation for this.  Small values of the
coefficient give rise to very narrow double peaks  at each cyclotron
energy in the spectral function.   This is the result of multiple
solutions to (\ref{eq:solve}), in fact there are two for each logarithmic
peak in $\Sigma^{'}(\epsilon)$.  The extreme narrowness of these
peaks suggests that these excitations are merely Landau level mixing.  This
conclusion is supported by the fact that for small values of $K_2$ the
gap is only very slightly reduced.  As $K_2$ is increased the double peaks
are reduced in size and eventually give rise to incoherent structures.
In the example we have been considering, $p=50$, this crossover
occurs when $K_2 \approx 0.5$ which is also where the gap
starts to decrease.
When the coefficient becomes large there is no longer any evidence
of simple Landau level mixing.

From the analysis it is clear that experimentalists should be rather 
cautious in fitting results for $\Delta \omega_c^{*}$ or
$m^{*}/m$ to SdH or longitudinal resistivity data - our numerical
results differ strongly from the analytic approximations
(\ref{eq:gapcf}) and (\ref{eq:gapcf2}) which have previously appeared
in the literature \cite{hlr,kim1,stern}.

However, the numerical results in Figs. \ref{fi:gap1cf} and \ref{fi:speccf}
are still based on the perturbative results in (\ref{eq:rselfecf2}) -
 (\ref{eq:ks}) for $\Sigma(\epsilon)$.   We now go a little beyond this
lowest-order perturbation theory framework.

\subsection*{4. Iterative Self-Consistent Results for $\Sigma(\epsilon)$ and
$\Delta\omega_c^{*}$}
As already explained in refs. \cite{theory}, the CF problem is
essentially a non-perturbative one.  The results in refs. \cite{theory}
address the $\nu = 1/2$ gapless system; an analysis of the $s=2$ case for
$\nu \neq 1/2$ was also given by Stern and Halperin \cite{stern}, but
taking no account of the structure described above.
 
Here we attempt to improve the results just given, by solving iteratively for
the self-energy.  We have found a way to do this which handles the
coherent pole contributions to $\Sigma(\epsilon)$ in an exact
rainbow summation; the incoherent parts must still be treated
approximately.

We start from the iterative equation
\begin{equation}
\Sigma^{(i+1)}(\epsilon) = \sum_m\int d\epsilon^{'} 
             D(\epsilon-\epsilon^{'})G^{(i)}_m(\epsilon^{'}) 
\label{eq:iterate}
\end{equation}
with
\begin{equation}
G^{(i)}_m(\epsilon) =  \frac{1}
     {\epsilon-(m+1/2)\Delta\omega_c+\Sigma^{(i)}(\epsilon)
-i\delta}.
\end{equation}
When the iterations converge we will have succeeded in summing over all
rainbow graphs. This kind of calculation is usually difficult but in
this case the analysis is simplified by the following.
First, because the real part
of $\Sigma^{(1)}(\epsilon)$ is bounded on either side of the gap 
$|\epsilon|<\Delta\omega_c/2$ by
logarithmic peaks, we are guaranteed that there are solutions
to (\ref{eq:solve}) for all $n$ within the gap. 
That the imaginary part of $\Sigma^{(1)}(\epsilon)$ is zero
in the gap implies that there are actually $2p$ $\delta$-function
peaks in $\sum_{m=-p}^{p-1}G^{(1)}_m(\epsilon)$
within the gap (exactly one for 
each $m$).  The distance 
between the peaks resulting from $m=0$ and $m=-1$ gives a renormalized gap,
$\Delta\omega_c^{(1)}$. 

Next we use $G^{(1)}$ in (\ref{eq:iterate}) to 
generate $\Sigma^{(2)}$.  We are particularly interested in the form of 
$\Sigma^{(2)}$ within the gap because this is where further corrections to the 
renormalized gap originate.  
The contributions to Im$G^{(1)}$ may be split into
two parts: a coherent piece, coming from the $\delta$-function peaks 
within the gap and an incoherent piece in the region outside the gap  (we 
use this terminology  for convenience- there may in fact
be solutions to (\ref{eq:solve}) in the latter region which 
correspond to poles in $G^{(0)}$).   Upon evaluating Re$\Sigma^{(2)}$ we
notice that the coherent parts of Im$G^{(1)}$ give
rise to exactly the same logarithmic divergences as seen in $\Sigma^{(1)}$, except
that the peaks are located at the poles 
of Im$G^{(1)}$ and each peak is  weighted by a factor
\begin{equation}
w(\epsilon_m) = \left|1+\frac{\partial\Sigma^{'}(\epsilon_m)}{\partial
	\epsilon}\right|^{-1}
\end{equation}
where $\epsilon_m$ are the positions of the poles.  However, in most
cases the weight coming from the poles within the gap is small 
compared to the weight of the incoherent part 
of Im$G^{(1)}_m$, which means that 
there may be significant contributions to Re$\Sigma^{(2)}$ that do 
not come from the poles
of Im$G^{(1)}_m$.  Only the coherent parts of Im$G^{(1)}$ contribute to 
Im$\Sigma^{(2)}(\epsilon)$ for $|\epsilon|<\Delta \omega_c/2$, which vanishes
within the renormalized gap, 
$|\epsilon| < \Delta\omega_c^{(1)}/2$.

The corrections
arising from coherent parts may be calculated 
exactly.  However we are forced to examine the
incoherent corrections using approximations, beginning with the use of
(\ref{eq:approx}).  This approximation has the effect of smearing out the
logarithmic peaks at the Landau levels.  As discussed above, these
features are a crucial element within the gap; however outside the
gap their role is not as important.  They may generate quasiparticles
peaks as solutions to (\ref{eq:solve}) but the weight of these peaks is
very small. The artificial 
smoothing of the Landau level structures would greatly
simplify a numerical
integration over these parts.  Instead of doing this, we make use of the
fact that the weight of the incoherent parts may be determined exactly,
since
$\int_{|\epsilon|>\Delta\omega_c/2}d\epsilon\mbox{Im}G^{(1)}_m(\epsilon) = 
1-w(\epsilon_m)$.   Therefore we approximate (\ref{eq:iterate}) as
\begin{equation}
\Sigma^{(2)}(\epsilon) = \sum_mD(\epsilon-\epsilon_{max})(1-w(\epsilon_m))
\end{equation}
where the peak of Im$G(\epsilon)$ is at $\epsilon_{max}$. 
(Note that there is still 
an implicit integration over $q$).  This approximation is valid for 
$\epsilon_{max}>\Delta\omega_c/2$, which occurs when 
$\frac{2K_2}{\pi}\log(2p+1) \approx 1$.  

\begin{figure}[h]
\epsfysize=3.6in
\epsfbox[22 140 561 715]{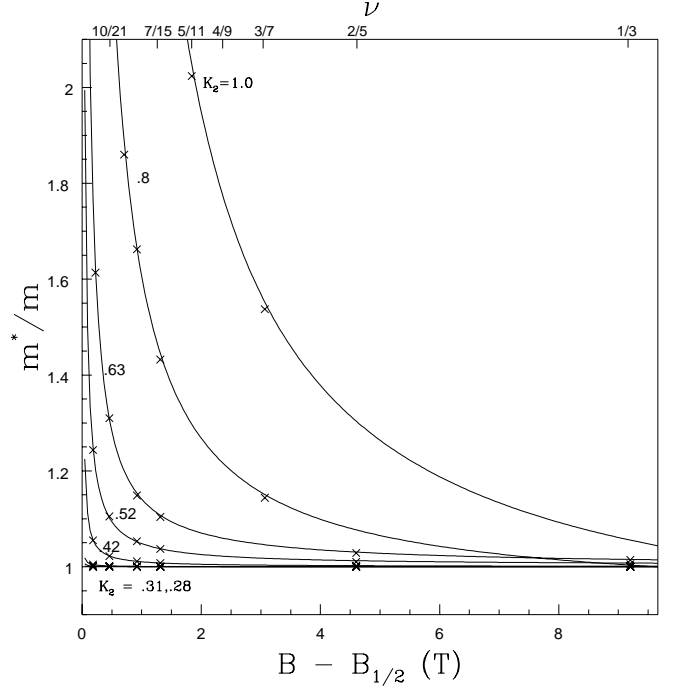}
\caption{The effective mass, $m^{*}$ as a function of $B$ and
$\nu$ for composite fermions. The lower $x-$ axis has been obtained
by assuming an electron density
$n_e = 2.23 \times 10^{11}\mbox{cm}^{-2}$.
The crosses are numerical calculations which include corrections coming
from $\Sigma^{(2)}$ (the self-consistent self-energy).  The curves are a
guide for the eye.\label{fi:gap2cf}}
\end{figure}

We have used this procedure 
to calculate the gap
as a function of $p$, keeping only results which appear to have converged
after the second step.  We find corrections 
result that are as large as 30\% of the first order result.
These calculations are shown in Fig. \ref{fi:gap2cf} 
plotted as $\frac{m^{*}}{m} = \frac{\Delta\omega_c}{\Delta\omega_c^{*}}$
for various couplings.   We estimate the actual coupling, 
in the experiment of Du et. al. \cite{du}  to be $K_2\approx 0.8$ 
(see the Appendix). 
We emphasize that the 
results for small values of $p$ are not meaningful because of
the approximation  (\ref{eq:matrix}), which assumes that $p$ is large,
thus there is no overlap with experimental results.
Therefore, we draw no conclusions about the exact relation between
$m^{*}$ and $p$.
However,  there is a range of $p$ where $p$ may be considered to be large
and  $\log(2p+1)$ is not. In this range we expect our results to be valid
but to have not yet reached the asymptotic limit of $m^{*}/m = K_2\log(2p+1)$.  
This calculation shows the effective mass to be far more sensitive
to the coupling than to $p$, thus it is difficult
determine what relation these results have to experiments without
knowing the coupling exactly. 

The information shown in Fig. \ref{fi:gap2cf} has been calculated using the
fillings shown on the upper axis for the different couplings shown on the plots.
This information may be interpreted in two ways.  First, we may assume that
the density, $n_e$, is fixed and that the different couplings arise by
variations of the other parameters in (\ref{eq:k2}). The values of $B$
shown in the lower $x$-axis have been determined using 
$n_e = 2.2\times 10^{11}\mbox{cm}^{-2}$.  This allows us to compare to 
the experimental data of Du. et. al \cite{du} - their data is consistent 
with a coupling of $K_2\approx 1$, which is slightly larger than the 
coupling $K_2 \approx 0.8$ estimated using (\ref{eq:k2}).  
Our calculations do not agree with  the results of Leadley et. al. 
\cite{leadl}, but their data does not go beyond $\nu = 3/7$, which
is not within the range of our approximations. (Obviously,
we cannot compare the results of Du et. al. at these values either.)

\begin{figure}[h]
\epsfysize=3.5in
\epsfbox[22 136 565 688]{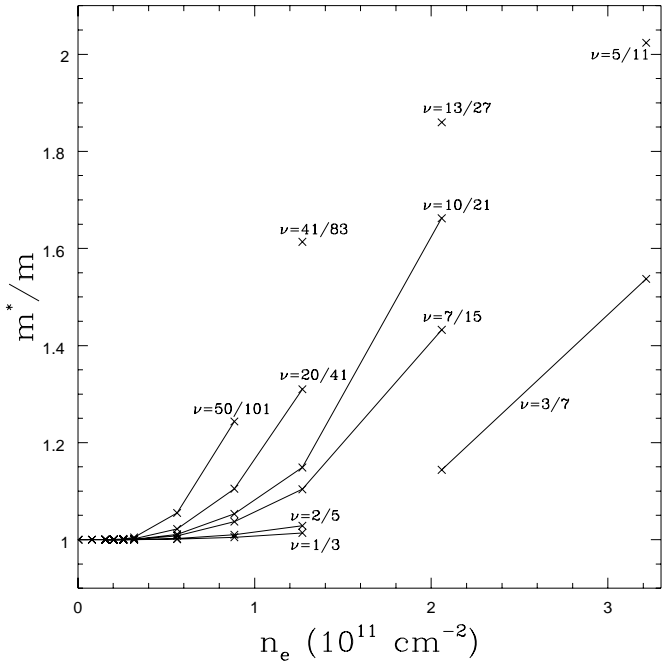}
\caption{The effective mass $m^{*}$ as a function of electron density.
These are the same results as in Fig. 10 except that the $x$-axis
has been obtained using
$n_e = \frac{1}{4\pi}\left(\frac{2 K_2 me^2}{\epsilon}\right)^2$ with
 $\tilde{\epsilon} = 13$ and $m= .07m_e$.
 The crosses are numerical
 calculations which include corrections coming
from $\Sigma^{(2)}$ (the self-consistent self-energy).
\label{fi:gap3cf}}
\end{figure}

Alternatively, we may assume that each curve in Fig. \ref{fi:gap2cf}
is associated with a different density, $n_e$.  Using 
$\tilde{\epsilon} = 13$ and $m=.07m_e$ in (\ref{eq:k2}) we replot the results
as shown in Fig. \ref{fi:gap3cf}.  The curves show that the effective mass
increases with electron density, which agrees with the observations
by Leadley et. al., but there is no quantitative agreement for the
range of data we have calculated, due to the fact our approximations
are not valid for the range of $\nu$ studied in the experiment. 

\subsection*{5. Summary}
The main point of this paper has been to deal directly with the singular
structure induced in $\Sigma(\epsilon)$ and $G(\epsilon)$ by Landau
level quantization, rather than smoothing over it, as has been
done it previous work.

As a result of this we find that the size of the CF gap in the 
gauge theory of the FQHE is considerably changed, particularly in the 
intermediate coupling regime which is experimentally relevant.   The CF
gauge theory is so far the only theory in which it seems feasible to 
calculate this gap (apart from finite-size numerical calculations),
for arbitrary stable filling fractions.   To test this theory it is 
necessary to take account of the singular structure we have discussed.  
At the present time gap measurements are subject to some
uncertainty, but our results indicate a significant discrepancy between 
claimed experimental results for the gap (or what is the same, 
$m^{*}/m$) and a self-consistent theory.

\begin{acknowledgements}
S. C. would like to acknowledge enlightening discussions with Professor 
Nathan Weiss, Igor Herbut and Martin Dub\'{e}.
\end{acknowledgements}


\appendix
\subsection*{Appendix}
In this appendix we give the details of the calculations of the
three self-energies considered in the main text.
\subsection*{(a) Debye Phonons}

In the self-energy calculations presented in the text we have
assumed a model similar to the one described by Ezawa et. al. \cite{ezawa}.  In 
this model it is assumed that phonons can propagate throughout
the semiconductor layers without reflection at the boundaries.  The 
electrons are confined to a very thin region which we approximate
to be two-dimensional. The usual forms of the electron-phonon 
interactions for both Debye and Einstein models may be used. 
For the Debye model, the electron-phonon interaction has the form
\begin{equation}
H_{int} = \Xi_D \int d^3r \psi_e^{\dag}(\vec{r})\psi_e(\vec{r})
          \nabla\cdot\vec{u}(r)
\end{equation}
where $\Xi_D$ is the deformation potential,
\begin{equation}
\vec{u}(\vec{r}) = \sum_{\vec{q}} \hat{e} 
     \sqrt{\frac{\hbar}{2\omega_{\vec{q}}\rho V}}
     (a_{\vec{q}}e^{i\vec{q}\cdot\vec{r}} + 
      a^{\dag}_{\vec{q}}e^{-i\vec{q}\cdot\vec{r}})
\end{equation}
is the displacement operator, with ion mass density $\rho$, cell volume $V=a^3$ 
and 
dispersion $\omega_{\vec{q}} = c_s \vec{q}$; and $\psi_e$ is the 
electron wavefunction expressed in a Landau level basis,
\begin{eqnarray}
\psi_e(\vec{r})& =& \sum_{n,k}c_{n,k}\left(\frac{1}{2^n n!l}\right)^{1/2}
          \left(\frac{m \omega_c}{\pi \hbar}\right)^{1/4}
       H_n(y/l-kl) \nonumber \\
& & \exp\left(-\frac{(y/l-kl)^2}{2}+ikx\right)(\delta(z))^{1/2}.
\label{eq:psi}
\end{eqnarray}
In these
expressions $\omega_c = eB/m$ is the cyclotron frequency, $m$ is the 
electron mass and 
$l = (\hbar/eB)^{1/2}$ is the magnetic length.
This leads to 
\begin{eqnarray}
H_{int}& =& \Xi_D\sum_{n,n^{'},k,k^{'},\vec{q}}
         \sqrt\frac{\hbar}{2\omega_{\vec{q}}\rho V}\hat{e}\cdot\vec{q}
         (a_{\vec{q}} + a_{-\vec{q}}^{\dag})c_{nk}^{\dag}c_{n^{'}k{'}}
\nonumber \\
  & &       \Lambda(n,n^{'},k,k^{'},\vec{q})
\end{eqnarray}
where
\begin{eqnarray}
& &\Lambda(n,n^{'},k,k^{'},\vec{q}) =\left(\frac{1}
                   {2^{n+n^{'}}n!n^{'}!}\right)^{1/2}\left(\frac{m\omega_c}
                      {\pi\hbar}\right)^{1/2}\frac{1}{l}
\nonumber \\
& & \hspace{.5in}
\times \int d^3r H_n(y/l-kl) H_{n^{'}}(y/l-k^{'}l)\delta(z)  \nonumber  \\
& &  \hspace{.5in} 
 \times \exp\left(-\frac{(y/l-kl)^2}{2}-\frac{(y/l-k^{'}l)^2}{2}+ iq_yy
     \right. \nonumber \\
& &     \hspace{1in} \left.  + i(-k+k^{'}+q_x)x +iq_zz \right)   \\
&& =  \frac{\delta(-k+k^{'}-q_x)}{l}
           \exp\left(-\frac{q^2l^2}{4}+iq_y(k+k^{'})l^2\right) \nonumber \\ 
& &      \left(\frac{2^{n^{'}}n!}{2^n n^{'}!}\right)^{1/2}
      \left(\frac{-q_x l-iq_yl}{2}\right)^{n^{'}-n}
      L^{n^{'}-n}_{n}\left(\frac{-q^2l^2}{2}\right) \label{eq:lambda}
\end{eqnarray}
for $n \leq n^{'}$ and $q^2 = q^2_x +q_y^2$. 
 The variables $k,k^{'}$ are of no concern as far as the 
properties of the electron Green's functions are concerned because the 
energy of the non-interacting
 electrons does not depend $k$ and the final form of
$H_{int}$ does not depend on $k,k^{'}$ or $q_z$. Furthermore, in what follows 
we assume that the phonon operator is isotropic
with respect to the $x$-$y$ plane.
In the limit of large $n\approx n^{'}$ and $ql\gg(n^{'}-n)/n$ the
 matrix element 
$|\Lambda|^2$ reduces to
\begin{equation}
|\Lambda(n,q)|^2 =  \frac{1}{\pi q} \left(\frac{eB}{2n}\right)^{1/2}
\end{equation} 
The final form of $H_{int}$ is
\begin{equation}
H_{int} =  \Xi_D \sum_{n,n^{'},q} \sqrt\frac{\hbar q}{2 c_s\rho V}
      (a_q + a_{-q}^{\dag})c_{n}^{\dag}c_{n^{'}}
      \Lambda(n,n^{'},q)
\end{equation}
 
The self-energy is derived from the second order contribution
of $H_{int}$
\begin{eqnarray}
\Sigma_n(\epsilon)& =& \Xi^2_D\left(\frac{\hbar}{2c_s\rho a}\right)
            \sum_{n^{'}=0}\int \frac{qd^2q}{(2\pi)^2} \int_0^{\omega_D}
             \frac{d\omega}{\pi}
          |\Lambda(n^{'},q)|^2 \nonumber \\
& &\times \frac{1}{\epsilon-\omega-(n+1/2)\omega_c-i\delta}
               \frac{1}{\omega-c_sq-i\delta}
\end{eqnarray}
where the factor $1/a$ comes from the $q_z$-integral.
We now shift the energies by an amount $p\omega_c$ which is the chemical 
potential when it lies halfway  between Landau levels.  The index is shifted
by $p$.  Performing the
integrals yields the  zero temperature results (\ref{eq:rselfephD}) and 
(\ref{eq:iselfephD}).  When the self-energy and the energy $\epsilon$ 
are expressed
in units of the Fermi energy, $E_F = p \omega_c$ 
the dimensionless coefficient is 
\begin{equation}
\bar{K_D} = \frac{\Xi^2_D m E_F}{4 \sqrt{2p} \pi c_s^3 \rho a} = 
      \frac{3\pi\Xi_D^2 E_F m}{2 \omega_D^3 \sqrt{4\pi n_e} aM}
       \label{eq:KD2},
\end{equation}
Using typical values for GaAs, ie.,
$\omega_D = (6\pi)^{1/3}c_s/a = 345$K, $\Xi_D = -4.4$eV, 
$m = .07m_e$, $M\approx10^5m_e$, $a=3.6\AA$,
$n_e = 2.25\times 10^{11}\mbox{cm}^{-2}$
  and $E_F \approx 1.5\times 10^{-3}$eV, one has $\bar{K_D}\approx 0.23$.
Equation (\ref{eq:KD2}) shows that there is no magnetic field dependence in
the coefficient.
The relation between $K_D$, which is given by 
(\ref{eq:KD}), and $\bar{K_D}$
is $K_D = \bar{K_D}/p$. The difference arises because
$K_D$ is used when the self-energy and the energy $\epsilon$
 are explicitly given in 
units of $\omega_c$, which is appropriate for solving (\ref{eq:solve}).

\subsection*{(b) Einstein phonons}
Following the approach of Engelsberg and Schreiffer we choose a coupling that
is independent of $q$
\begin{equation}
H_{int}  = g \sum_{n,n^{'},q}(a_q + a_{-q}^{\dag})c_n^{\dag}c_{n'}
          \Lambda(n,n^{'},q)
\end{equation}
This leads to a self-energy
\begin{eqnarray}
\Sigma_n(\epsilon)& =&  g^2 \sum_{n^{'}}|\Lambda(n,n^{'},q)|^2 a^2
             \int \frac{d^2q}{(2\pi)^2} \int_0^{\infty}\frac{d \omega}{\pi}
\nonumber \\
           &\times &  \frac{1}{\epsilon-\omega-(n+1/2)\omega_c -i\delta}
             \frac{1}{\omega-\omega_E-i\delta}
\end{eqnarray}
where we have assumed that 
the normalization of the $q$-integral is $(2\pi)^2/a^2$.
Equations (\ref{eq:rselfephE}) and (\ref{eq:iselfephE}) follow directly;
with $\Sigma$ and the energy $\epsilon$
expressed in units of $E_F$, the dimensionless coefficient
is 
\begin{equation}
\bar{K_E} = \frac{g^2 (2p)^{1/2} a}{4 E_F^2 l} = 
       \frac{g^2 a (4\pi n_e)^{1/2}}{4 E_F^2}
\end{equation}
In this form there is no dependence of the coupling on the magnetic 
field.

\subsection*{(c) Gauge Fluctuations} 
The original theory of strongly interacting electrons in a magnetic 
field takes the form
\begin{eqnarray}
{\cal L}& =&  \psi_e^{\dag}(x)(-i\partial_t)\psi_e(x)
+ \psi_e^{\dag}(x)\frac{(i\partial_i - eA_i)^2}{2m}\psi_e(x)
\nonumber \\
  & &     + \int d^2y \psi_e^{\dag}(x)\psi_e(x)V(x-y)\psi_e^{\dag}(y)\psi_e(y)
\end{eqnarray}
where $B\equiv \nabla \times A$ is the external magnetic field.

The electrons are transformed into composite fermions by attaching
two magnetic flux quanta to each.   The constraint that this additional magnetic
field, $b$, is proportional to the electron density is 
\begin{equation}
b \equiv \nabla \times a = 4\pi \psi^{\dag}(x)\psi(x)
\label{eq:constraint}
\end{equation}
 and is implemented
with the use of a Chern-Simons term in $\cal{L}$.   In addition, the
CF's experience a gauge field that is the difference between the 
external field $A$ and the ``statistical" field $a$, $A-a= \Delta A$:
\begin{eqnarray}
{\cal L} &=& \psi^{\dag}(x)(-i\partial_t+a_0)\psi(x)
          + \psi^{\dag}(x)\frac{(i\partial_i - e\Delta A_i)^2}{2m}\psi(x)
\nonumber \\
& &          +\frac{a_0\nabla\times a}{4\pi} 
     +\int d^2 y \frac{\nabla\times a(x)V(x-y)\nabla\times a(y)}{16\pi^2}
\end{eqnarray}
where the third term is the Chern-Simons term and 
we have used the constraint (\ref{eq:constraint})
to rewrite the last term. In this expression
$\psi$ is a fermionic operator representing composite fermions.
The theory is
completed by allowing fluctuations of the gauge field, $\delta a$,
\begin{eqnarray}
{\cal L} &=& \psi^{\dag}(x)(-i\partial_t+a_0)\psi(x) \nonumber \\
& &
       + \psi^{\dag}(x)\frac{(i\partial_i - e\Delta A_i-\delta a)^2}{2m}\psi(x)
      + \delta a_iD^{-1}_{ij}\delta a_j
\end{eqnarray}
The last term is the effective action of the gauge fluctuations,
which yields (\ref{eq:gauge}) \cite{lopez},
\cite{hlr}.  The second term yields the free
CF Hamiltonian,
\begin{equation}
H_{CF} = \frac{\psi^{\dag}(x)(i\partial_i -e\Delta A_i)^2\psi(x)}{2m}
\end{equation}
and the CF-gauge fluctuation interaction,
\begin{equation}
H_{int} = \frac{\delta a_i(x)\psi^{\dag}(x)(i\partial_i -e\Delta A_i)\psi(x)}
          {m}            
\end{equation}
in the Coulomb gauge, $q\cdot\delta a = 0$.
The eigenfunctions of $H_{CF}$ are the same as (\ref{eq:psi}) except 
that $\omega_c$ is replaced by $\Delta \omega_c$.
The operator $V_i \equiv (i\partial_i -e\Delta A_i)/m$ introduces additional
complications in the determination of the final form of $H_{int}$.
We are only considering interactions with the transverse component of 
$\delta a = \delta a_1$, corresponding to the transverse component of 
$V_1 = \cos \theta V_x + sin\theta V_y$. $V_x$ and $V_y$ can both be 
expressed in terms of creation and annihilation operators which act on the 
harmonic oscillator part of $\psi$:
\begin{eqnarray}
V_x & = & \sqrt{\frac{\Delta \omega_c}{2m}}(a+a^{\dag}) \\
V_y & = & i\sqrt{\frac{\Delta \omega_c}{2m}}(a-a^{\dag}) \\
V_1 & = & \sqrt{\frac{\Delta \omega_c}{2m}}(a e^{i\theta}+a^{\dag}e^{-i\theta})
\end{eqnarray}
In this case the matrix element will be 
\begin{eqnarray}
\Lambda_{CF}(n,n^{'},q)& =& \sqrt{\frac{\Delta \omega_c}{2m}}(
           \sqrt{n^{'}}\Lambda(n,n^{'}-1,q)e^{i\theta}\nonumber \\
& & + \sqrt{n^{'}+1}\Lambda(n,n^{'}+1,q)e^{-i\theta})
\end{eqnarray}
which, in the limit of large $n\approx n^{'}=p$, yields
\begin{equation}
|\Lambda_{CF}(n,n^{'},q)|^2 \approx \frac{4p\Delta\omega_c}{2m}|\Lambda(n,q)|^2.
\end{equation} 
The coefficient is 
\begin{equation}
   \frac{4p\Delta\omega_c}{2m}\approx \frac{2E_F}{m}
\end{equation}
Thus we are left with
\begin{eqnarray}
& &\Sigma_n(\epsilon) = \frac{2E_F}{m}\int\frac{d^2q}{(2\pi)^2}
          \int_0^{\infty}\frac{d \omega}{\pi}
          \int_{-\infty}^{\infty}\frac{d\epsilon^{'}}{\pi}\sum_{n^{'}=0}
    |\Lambda(n,n^{'},q)|^2 \nonumber \\
& &\mbox{Im}D(q,\omega)\mbox{Im}G_{n^{'}}^0(\epsilon^{'})
   \left(\frac{\theta(\epsilon^{'})}{\epsilon-\epsilon^{'}-\omega+i\delta}
        +\frac{\theta(-\epsilon^{'})}{\epsilon - \epsilon^{'}+\omega+i\delta}
   \right) \nonumber \\
& &
\end{eqnarray}
Just as in the phonon case, a cut-off in the sum at $n^{'} = 2p$ is needed.
Equations (\ref{eq:rselfecf2}), (\ref{eq:iselfecf2}), (\ref{eq:rselfecfs})
and (\ref{eq:iselfecfs}) follow upon integration.  When the self-energy and 
the energy $\epsilon$ are 
expressed in units of the Fermi energy, $E_F = p\Delta\omega_c$,  the
dimensionless coupling is
\begin{equation}
K_2 = \frac{E_F \epsilon}{e^2}\left(\frac{1}{4\pi n_e}\right)^{1/2}
    \approx \left(\frac{n_0}{n_e}\right)^{1/2},
\end{equation}
where $n_0 = \frac{E_F\tilde{\epsilon}}{e^2}\sqrt{\frac{1}{4\pi}}
\approx 1.5 \times 10^{11} \mbox{cm}^{-2}$ is the density at which 
$K_2 = 1$.  For the experiment of Du et. al. \cite{du}, one has 
 $E_F = 2\pi n_e/m$,$n_e = 2.3\times10^{11}\mbox{cm}^{-2}$,
$m=.07m_e$ and $\tilde{\epsilon} = 13$, so one expects $K_2 \approx 0.8$.
Note that equations (27)-(29) assume that the self-energy has units
of $\Delta \omega_c$, which is more appropriate for solving the 
self-consistent equation, (\ref{eq:solve}).

\end{document}